\documentclass[final]{svjour3}

\usepackage{graphicx}

\usepackage{rotating}
\usepackage{multirow}
\usepackage{amssymb}
\usepackage{mathptmx}
\usepackage{siunitx}
\usepackage{hyperref}
\usepackage[square,numbers]{natbib}
\usepackage{xcolor}
\newcolumntype{C}{>{\centering\arraybackslash}p{5em}}

\makeatletter
\journalname{Journal of Low Temperature Physics}

\begin{document}

\newcommand{\hdblarrow}{H\makebox[0.9ex][l]{$\downdownarrows$}-}

\title{In--flight performance of the LEKIDs of the OLIMPO experiment}

\authorrunning{A. Paiella et al.}
\titlerunning{In--flight performance of the LEKIDs of the OLIMPO experiment}

\author{A. Paiella$^{1,2,*}$ \and P.~A.~R. Ade$^{3}$ \and E.~S. Battistelli$^{1,2}$\and  \\M.~G. Castellano$^{4}$ \and I. Colantoni$^{5}$ \and F. Columbro$^{1,2}$\and A. Coppolecchia$^{1.2}$ \and G. D'Alessandro$^{1,2}$ \and \\P. de Bernardis$^{1,2}$  \and M. De Petris$^{1,2}$ \and S. Gordon$^{6}$ \and L. Lamagna$^{1,2}$ \and C. Magneville$^{7}$\and S. Masi$^{1,2}$\and \\P. Mauskopf$^{6,8}$\and G. Pettinari$^{4}$\and F. Piacentini$^{1,2}$\and G. Pisano$^{3}$\and G. Polenta$^{9}$\and G. Presta$^{1,2}$\and \\E. Tommasi$^{9}$\and C. Tucker$^{3}$\and V. Vdovin$^{10,11}$\and \\A. Volpe$^{9}$\and D. Yvon$^{7}$}

\institute{$^{1}$ Dipartimento di Fisica, \emph{Sapienza} Universit\`{a} di Roma, P.le A. Moro 2, 00185 Roma, Italy
\at$^{2}$ Istituto Nazionale di Fisica Nucleare, Sezione di Roma, P.le A. Moro 2, 00185 Roma, Italy
\at$^{3}$ School of Physics and Astronomy, Cardiff University, Cardiff CF24 3YB, UK
\at$^{4}$ Istituto di Fotonica e Nanotecnologie--CNR, Via Cineto Romano 42, 00156 Roma, Italy
\at$^{5}$ CNR--Nanotec, Institute of Nanotechnology, \\c/o Dipartimento di Fisica, \emph{Sapienza} Universit\`{a} di Roma, P.le A. Moro 2, 00185, Roma, Italy
\at$^{6}$ School of Earth and Space Exploration, Arizona State University, Tempe, AZ 85287, U.S.A
\at$^{7}$ IRFU, CEA, Universit\`{e} Paris--Saclay, F--91191 Gif sur Yvette, France
\at$^{8}$ Department of Physics, Arizona State University, Tempe, AZ 85257, U.S.A
\at$^{9}$ Italian Space Agency, Roma, Italy
\at$^{10}$ Institute of Applied Physics RAS, State Technical University, Nizhny Novgorov, Russia
\at$^{11}$ ASC Lebedev PI RAS, Moscow, Russia
\\$^{*}$ \email{alessandro.paiella@roma1.infn.it}}

\maketitle

\begin{abstract}

We describe the in--flight performance of the horn--coupled Lumped Element Kinetic Inductance Detector arrays of the balloon--borne OLIMPO experiment. These arrays have been designed to match the spectral bands of OLIMPO: 150, 250, 350, and \SI{460}{\giga\hertz}, and they have been operated at \SI{0.3}{\kelvin} and at an altitude of \SI{37.8}{\kilo\metre} during the stratospheric flight of the OLIMPO payload, in Summer 2018. During the first hours of flight, we tuned the detectors and verified their large dynamics under the radiative background variations due to elevation increase of the telescope and to the insertion of the plug--in room--temperature differential Fourier transform spectrometer into the optical chain. We have found that the detector noise equivalent powers are close to be photon--noise limited and lower than those measured on the ground. Moreover, the data contamination due to primary cosmic rays hitting the arrays is less than 3\% for all the pixels of all the arrays, and less than 1\% for most of the pixels. These results can be considered the first step of KID technology validation in a representative space environment.


\end{abstract}

\section{Introduction}

Modern Precision Cosmology measurements require kilo--pixel arrays of fast and ultra--sensitive cryogenic detectors, working in a large spectral band, from 60 to \SI{600}{\giga\hertz} \cite{CORE_1}. Kinetic Inductance Detectors (KID) \cite{KIDs} fulfill all these requirements together with the simplicity in the micro--fabrication process and in the cold bias/readout electronics. 

KIDs are low-temperature supercontuctive resonators, intrinsically multiplexable in the frequency domain, and with response time of the order of tens of microseconds. Lumped element KIDs (LEKIDs) \cite{Doyle} are obtained by properly shaping and sizing a superconductive film on a dielectric substrate in order to use the inductor also as a radiation absorber for the wavelengths of interest. KIDs are pair-breaking detectors, where incident photons with energy larger than the Cooper pair binding energy of the superconductor can be absorbed and transduced by the resonator as a change in its resonant frequency and quality factor. Thanks to their properties, KIDs can be used in a wide electromagnetic wavelength range.

In the microwaves, KID technology has been already proven for ground--based experiments by NIKA \cite{NIKA} and NIKA2 \cite{NIKA2}. OLIMPO \cite{Coppolecchia} has qualified the KID technology in a representative near--space
environment, relevant for TRL (Technology Readiness Level) advancement in view of future space missions \cite{CORE_1,CORE_2,CORE_3,CORE_4,CORE_5,CORE_6,CORE_7,CORE_8,CORE_9,CORE_10}. In the past years, several efforts have been made to characterize KIDs in representative space conditions \cite{monfardini, Baselmans, Karatsu}. During the 2018 flight, OLIMPO operated four arrays of LEKIDs in the stratosphere at an altitude of \SI{37.8}{\kilo\metre}. 

The OLIMPO experiment (\url{http://olimpo.roma1.infn.it/}) is a millimeter--wave observatory with a \SI{2.6}{\metre} aperture telescope devoted to the measurement of the spectrum of the Sunyaev Zel'dovich (SZ) effect \cite{SZ} in clusters of galaxies. The approach consists in using a plug--in room--temperature differential Fourier Transform spectrometer (DFTS) \cite{schillaci2014} coupled to four detector arrays cooled at \SI{0.3}{\kelvin} by a $^{3}$He refrigerator in a wet liquid nitrogen and liquid helium cryostat \cite{Coppolecchia_cryo}. 
Bandpass filters and dichroichs select the individual array loading and frequency range which illuminate each of the detector arrays. 
Center frequencies are 150, 250, 350, \SI{460}{\giga\hertz} and 17\%, 36\%, 9\%, 13\% bandwidths respectively.
With this design, low resolution spectroscopy ($\Delta\nu=\SI{5}{\giga\hertz}$) in the four photometric bands can be performed to allow degeneracy breaking in the determination of the intracluster gas parameters \cite{deBernardis}.

\section{Detector design and simulations}
\label{sec:detector}

The design, simulation and fabrication of the detector arrays of OLIMPO are described in detail in \cite{Paiella_isec,Paiella_ground}. Here, we report a brief summary. 

The OLIMPO detector system consists of horn-coupled LEKIDs working at 150, 250, 350, and \SI{460}{\giga\hertz}. The detector system, including the waveguide, the transition element (if needed), the absorber, the dielectric substrate and the backshort, was optimized for each band through optical simulations performed with the ANSYS HFSS \cite{HFSS} software. 

The material and thickness of the superconducting film were chosen as trade--off between critical temperature and kinetic inductance. In addition, the OLIMPO detectors had to be designed to operate at \SI{0.3}{\kelvin} and for a minimum detectable frequency of \SI{135}{\giga\hertz}. These constraints determined the choice of aluminum \SI{30}{\nano\metre} thick, for which we measured the critical temperature, $T_{c}=\SI{1.31}{\kelvin}$, the residual resistance ratio, ${\rm RRR}=3.1$, the sheet resistance, $R_s=\SI{1.21}{\ohm/\Box}$, and the surface inductance, $L_{s}=\SI{1.38}{\pico\henry/\Box}$, \cite{Paiella_Wband, Paiella_wolte}.

\begin{figure}[htbp]
\vspace{-0.2cm}
\begin{center}
\includegraphics[scale=0.41]{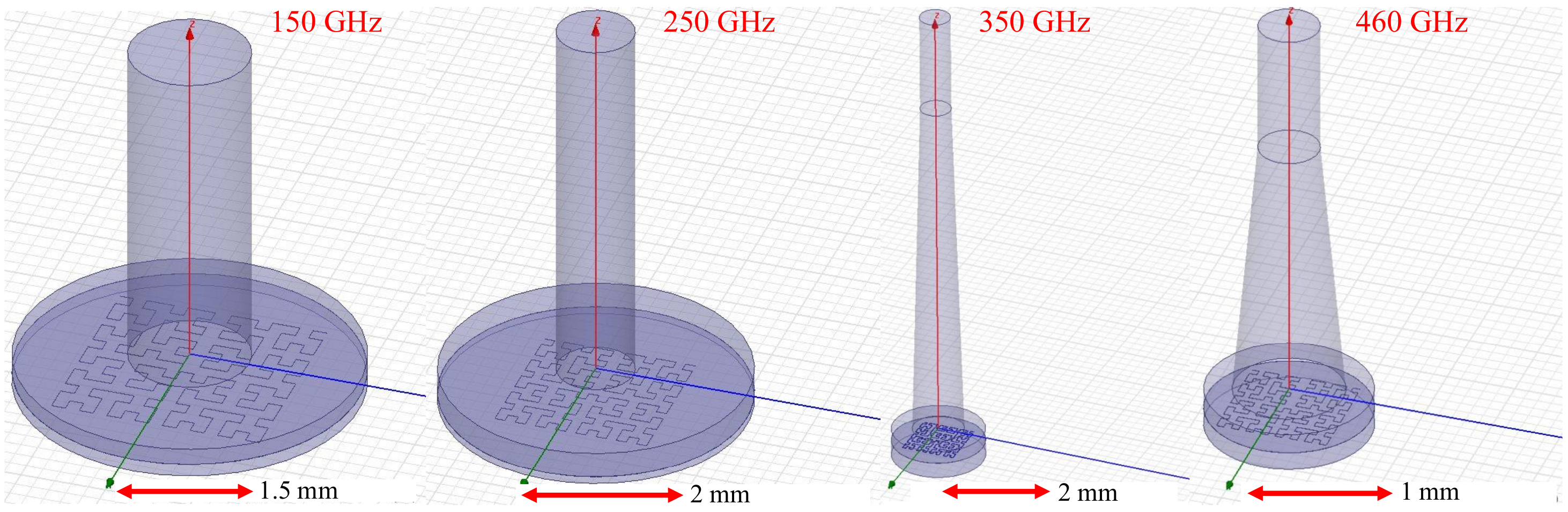}
\caption{HFSS design of the four detector systems of the OLIMPO experiment. From {\it left} to {\it right}: the \SI{150}{\giga\hertz}, \SI{250}{\giga\hertz}, \SI{350}{\giga\hertz}, and \SI{460}{\giga\hertz} systems. For the \SI{150}{\giga\hertz} and the \SI{250}{\giga\hertz} systems, from {\it top} to {\it bottom}: the circular waveguide, the free--space, the absorber, the dielectric substrate and the backshort. For the \SI{350}{\giga\hertz} and the \SI{460}{\giga\hertz} systems, from {\it top} to {\it bottom}: the circular waveguide, the flare, the free--space, the absorber, the dielectric substrate and the backshort. Drawings are not in scale.}
\label{fig:HFSS}
\end{center}
\vspace{-0.2cm}
\end{figure}

Fig.~\ref{fig:HFSS} shows the HFSS design of the optimized detector systems for the four bands of the OLIMPO experiment: front--illuminated IV order Hilbert patterns, where the characteristic length scales with the observed radiation frequency; the \SI{150}{\giga\hertz} and \SI{250}{\giga\hertz} arrays are coupled to the radiation via a single--mode circular waveguide, while the \SI{350}{\giga\hertz} and \SI{460}{\giga\hertz} are coupled via a single--mode flared circular waveguide. The dielectric substrate is made of silicon with thickness depending on the observed radiation frequency. The back surface of the substrate (the one opposite to the detectors one) is metallized with aluminum \SI{200}{\nano\metre} thick in order to act as a backshort and as a phonon trap.

These simulations aimed at maximizing the power absorbed by the patterned surface and minimizing the losses through the lateral surfaces of the cylinders schematizing the substrate and the free--space.

Tab.~\ref{tab:param_geom} collects the values of the OLIMPO receiver sizes optimized through the optical simulations: substrate thickness, $t_{s}$, backshort distance, $d$, Hilbert characteristic length, $s_h$, and width, $w_{h}$, waveguide diameter, $d_{wg}$, and height, $h_{wg}$, flare aperture, $d_{f}$, and height, $h_{f}$. 

\begin{table}[htbp]
\vspace{-0.2cm}
\centering
\fontsize{0.28cm}{0.45cm}\selectfont{
		\begin{tabular}{c|c|c|c|c|c|c|c|c}
		\hline
		\hline
		\multicolumn{1}{c|}{\multirow{1}{*}{Channel}}&
		\multicolumn{1}{c|}{\multirow{1}{*}{$t_{s}$}}&
		\multicolumn{1}{c|}{\multirow{1}{*}{$d$}}&
		\multicolumn{2}{c|}{\multirow{1}{*}{Absorber}}&
		\multicolumn{2}{c|}{\multirow{1}{*}{Waveguide}}&
		\multicolumn{2}{c}{\multirow{1}{*}{Flare}}\\
		\cline{4-9}
		\multicolumn{1}{c|}{\multirow{1}{*}{$\left[\SI{}{\giga\hertz}\right]$}}&
		\multicolumn{1}{c|}{\multirow{1}{*}{$\left[\SI{}{\micro\metre}\right]$}}&				\multicolumn{1}{c|}{\multirow{1}{*}{$\left[\SI{}{\micro\metre}\right]$}}&
		\multicolumn{1}{c|}{\multirow{1}{*}{$s_{h}$ $\left[\SI{}{\micro\metre}\right]$}}&
		\multicolumn{1}{c|}{\multirow{1}{*}{$w_{h}$ $\left[\SI{}{\micro\metre}\right]$}}&
		\multicolumn{1}{c|}{\multirow{1}{*}{$d_{wg}$ $\left[\SI{}{\milli\metre}\right]$}}&
		\multicolumn{1}{c|}{\multirow{1}{*}{$h_{wg}$ $\left[\SI{}{\milli\metre}\right]$}}&
		\multicolumn{1}{c|}{\multirow{1}{*}{$d_{f}$ $\left[\SI{}{\milli\metre}\right]$}}&
		\multicolumn{1}{c}{\multirow{1}{*}{$h_{f}$ $\left[\SI{}{\milli\metre}\right]$}}\\
		\hline
		\hline
150&135&450&162&2&1.4&6&&\\
250&100&350&132&2&1.0&5&&\\
350&310&250&72&2&0.60&2&1.0&7\\
460&135&150&52&2&0.44&2&0.8&2\\
\hline
\hline
\end{tabular}		
}
\vspace{-0.cm}
\caption{Values of the parameters optimized through the optical simulations.}
\phantomsection\label{tab:param_geom}
\vspace{-0.2cm}
\end{table}

The size of the aberration--corrected focal plane in OLIMPO is such that the \SI{150}{\giga\hertz} and \SI{250}{\giga\hertz} array are hosted on a \SI{3}{inches} wafer, and the \SI{350}{\giga\hertz} and \SI{460}{\giga\hertz} array on a \SI{2}{inches} wafer. The aperture of the horns are such that the allowed number of active pixels is 19, 37, 23 and 41 for the 150, 250, 350 and \SI{460}{\giga\hertz} array, respectively. To these we added 4 dark pixels on the \SI{150}{\giga\hertz} array and 2 for each remaining array.

The capacitors were designed to have resonant frequencies in the [100;$\;$600]~MHz range, in order to satisfy the lumped element condition, and to reduce the TLS (two--level system) noise \cite{TLS}. Moreover, the resonant frequencies were chosen to couple two arrays to the same bias/readout line: the \SI{150}{\giga\hertz} with the \SI{460}{\giga\hertz} array, and the \SI{250}{\giga\hertz} with the \SI{350}{\giga\hertz} array. The coupling between the resonators and the \SI{50}{\ohm}--matched microstrip feedline is performed by means of capacitors, designed to have $Q_{c}\sim\SI{15000}{}$, guaranteeing a large detector dynamics, without substantially degrading the sensitivity. The resonant frequencies, the lumped condition, the values of $Q_{c}$ and the impedance of the feedline are simulated with the SONNET software \cite{SONNET}.

\begin{figure}[htbp]
\vspace{-0.4cm}
\begin{center}
\includegraphics[scale=0.43]{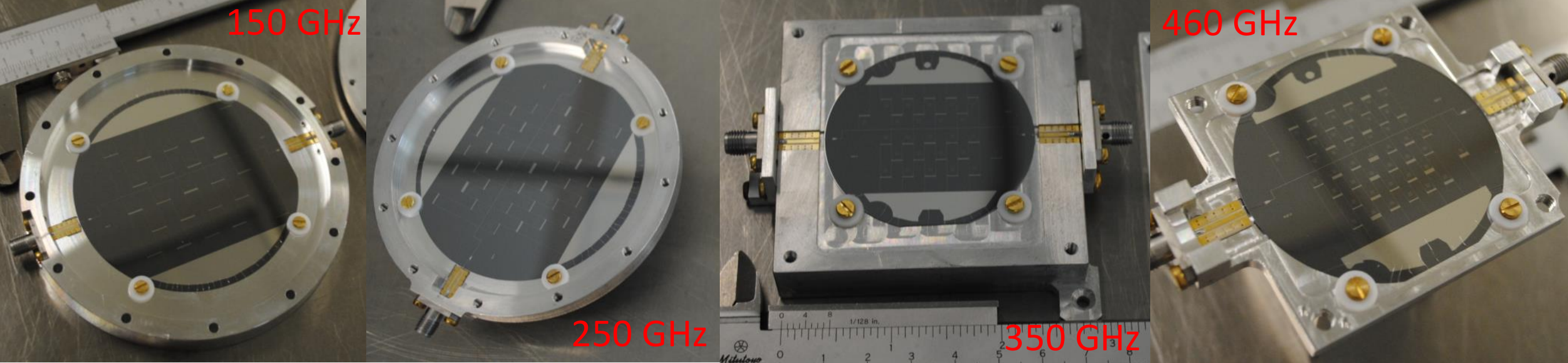}
\caption{Pictures of the four OLIMPO detector arrays mounted in their ergal (aluminum 7075) holders by means of four Teflon
washers \SI{100}{\micro\metre} thick. From {\it left} to {\it right}: the \SI{150}{\giga\hertz}, \SI{250}{\giga\hertz}, \SI{350}{\giga\hertz}, and \SI{460}{\giga\hertz} arrays.}
\label{fig:photo}
\end{center}
\vspace{-0.4cm}
\end{figure}

Fig.~\ref{fig:photo} collects the pictures of the four OLIMPO arrays, anchored in their holders through four Teflon washers $\SI{100}{\micro\metre}$ thick.

The performance of the OLIMPO detector arrays at ground are reported in \cite{Paiella_ground}.

\section{In--flight readout system}
\label{sec:readout_system}
The bias/readout system is composed of two independent chains, each serving two detector arrays and based on one ROACH2 board, including a MUSIC DAC/ADC board, and analog microwave components as amplifiers, bias tees, IQ modulator, IQ demodulator and attenuators. These systems, together with the firmware and the client software, were developed, tuned and characterized on the ground and described in \cite{Paiella_wolte, Gordon}.

\begin{figure}[htbp]

\begin{center}
\includegraphics[scale=0.5]{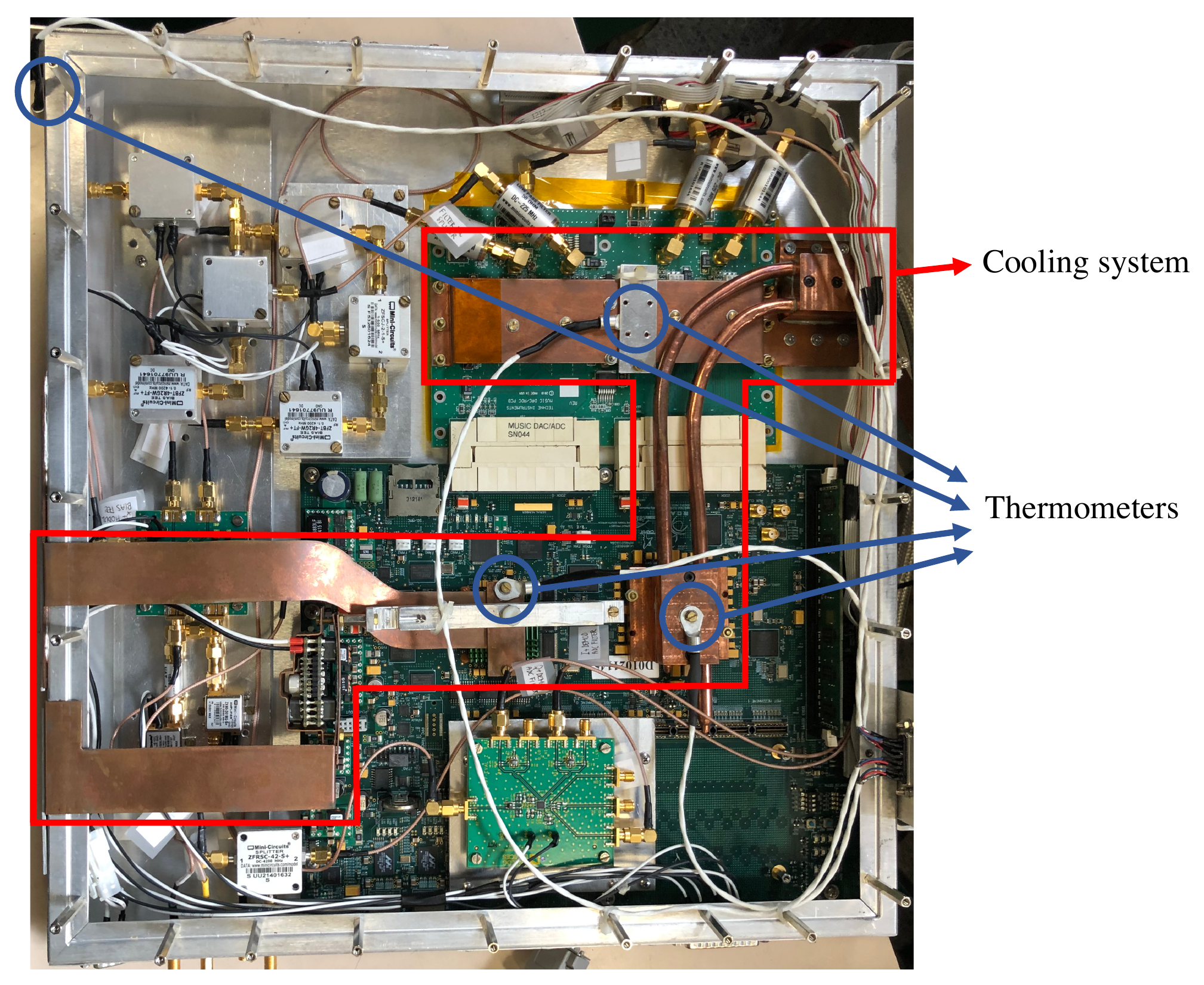}
\caption{Picture of the bias/readout electronics modified to work in the stratosphere. The {\it red box} highlights the cooling system composed of copper thermal straps and heat pipes, and the {\it blue circles} highlight the thermometers (MAXIM DS18B20) used to monitor the temperatures.}
\label{fig:ROACH}
\end{center}
\vspace{-0.2cm}
\end{figure}

The power dissipation system of the bias/readout electronics was modified in order to allow the components to work nominally in the stratospheric environment, where the external pressure of about \SI{3}{\milli\bar} results in a drastic reduction of convective cooling with respect to the ground. We built a system of copper straps and heat pipes, see fig.~\ref{fig:ROACH}, which thermally link the highest dissipation components to a radiator, maintaining their around $\SI{60}{\degreeCelsius}$ at float during the flight. The temperature of the most critical parts was monitored with MAXIM DS18B20 temperature sensors. The ROACH2 with the whole dissipation system has been successfully tested in a thermal--vacuum chamber (Weiss--WK1--1500/70).

\section{In--flight performance}
\label{sec:performance}
\begin{figure}[htbp]
\begin{center}
\includegraphics[scale=0.265]{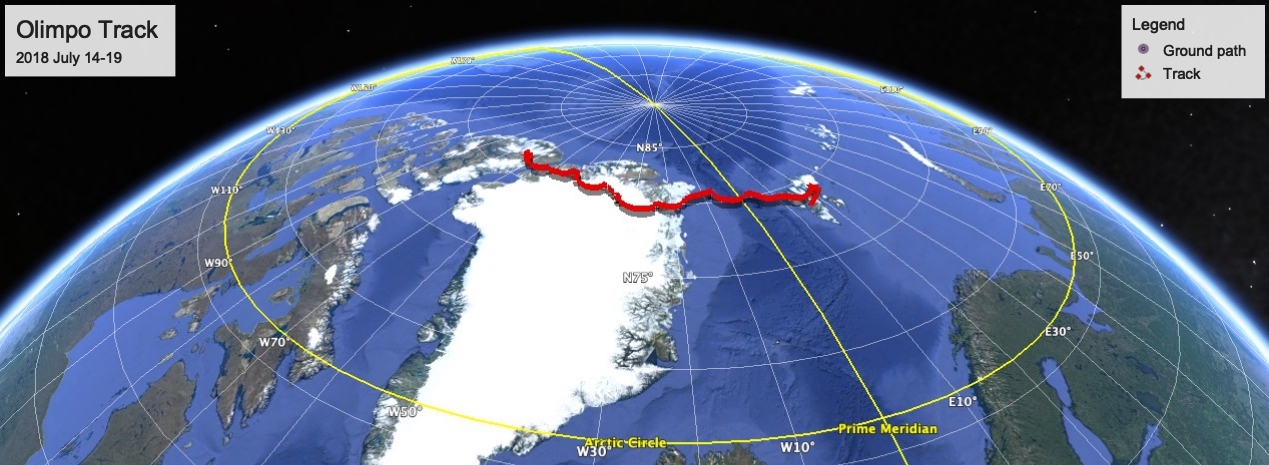}
\caption{Ground path and trajectory of the OLIMPO payload during the flight.}
\label{fig:traj}
\end{center}
\end{figure}

OLIMPO was launched from the Longyearbyen airport  ($\SI{78}{\degree N}$), in Svalbard Islands on July 14$^{\rm th}$ 2018, at 07:07 GMT. It reached the altitude of \SI{37.8}{\kilo\metre} after about \SI{5.5}{hours} and floated along the trajectory shown in fig.~\ref{fig:traj} for about \SI{5}{days} \cite{Masi_balloon}. 

\begin{figure}[htbp]
\vspace{-0.2cm}
\begin{center}
\includegraphics[scale=0.021]{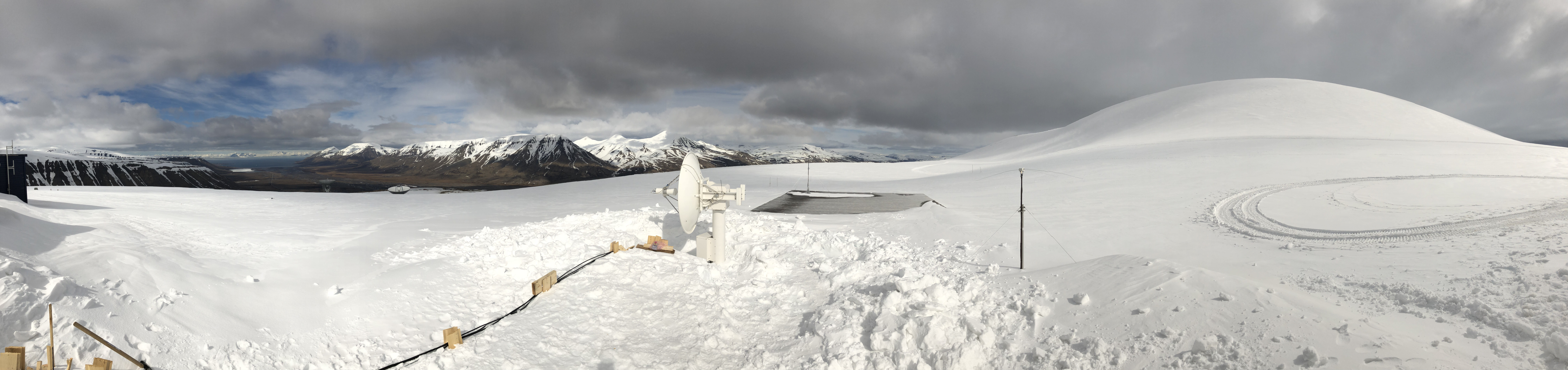}
\caption{Panoramic picture of the Longyearbyen fiord seen from the Kjell Henriksen Observatory (KHO) where the line--of--sight (LOS) telemetry was installed. The antenna of the LOS telemetry is shown in the picture.}
\label{fig:photo_LOS}
\end{center}
\end{figure}

All the measurements described in this paper were carried out in the first day of the flight, when the fast  (\SI{500}{kbps}) bidirectional line–-of–-sight (LOS) telemetry (L--band, IRILASI \& ELTA--ECA Group) was active, see fig.~\ref{fig:photo_LOS}.

The detailed analysis of the in--flight tuning, performance and impact of cosmic rays of the OLIMPO detectors is described in \cite{Masi_inflight}.

\subsection{Tuning, Calibration Lamp and Noise Equivalent Power}
\label{subsec:lamp}
\begin{figure}[htbp]
\vspace{-0.2cm}
\begin{center}
\includegraphics[scale=0.29]{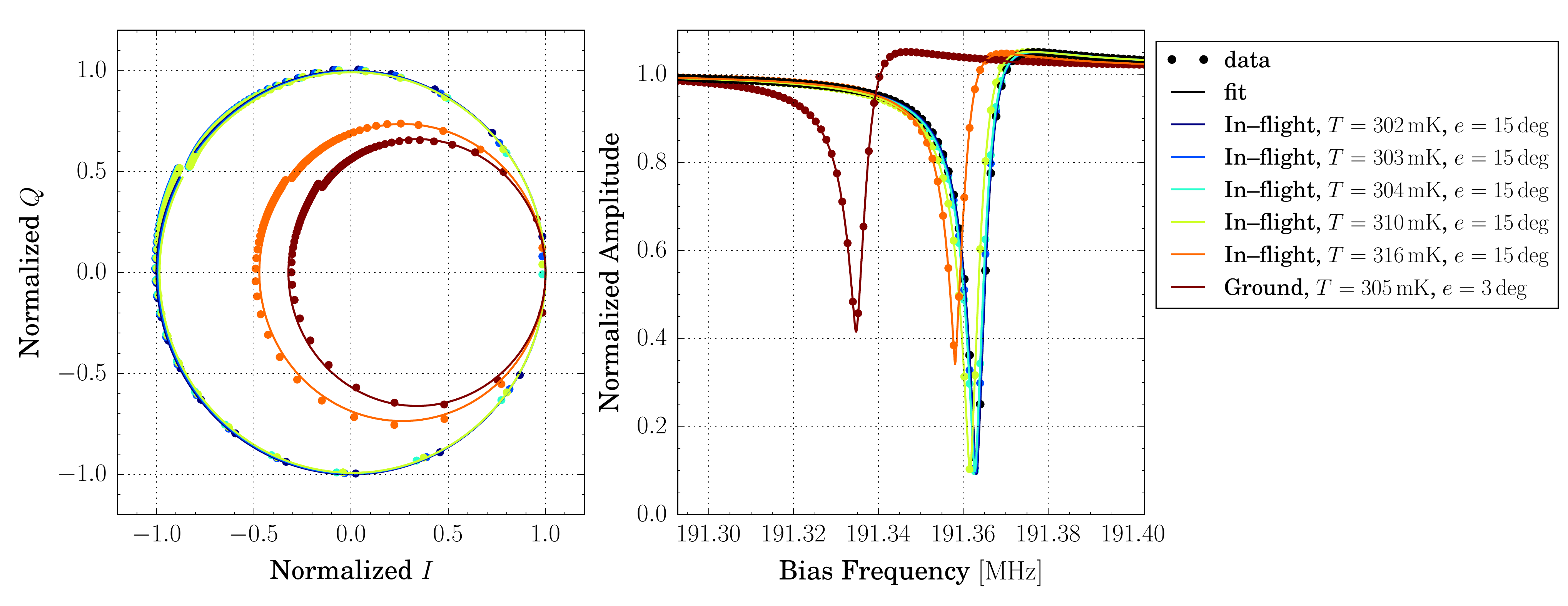}
\caption{Complex ({\it left}) and amplitude ({\it right}) $S_{21}$ parameter for pixel \#6 of the \SI{150}{\giga\hertz} array. The {\it dots} are the measured data and the {\it solid lines} are the best fits. In--flight and ground measurements are compared as described in the legend.}
\label{fig:tuning}
\end{center}
\vspace{-0.2cm}
\end{figure}

The tuning of all the detectors was performed during the first hour after reaching the float altitude, thanks to the fast bidirectional connection guaranteed by the LOS telemetry. As an example, fig.~\ref{fig:tuning} shows five tuning results performed while the refrigerator recovered its temperature after the shock of the launch, compared to the best tuning obtained on the ground. As expected, the resonant frequency moves to a lower value when the working temperature increases as well as the background increases, like in the ground operation. 

During the flight, we were able to monitor the performance of the detectors under background changes (due to changes of the telescope boresight elevations and the insertion of the DFTS) by means of a calibration lamp \cite{lamp_cal}. It is composed of a resistor heating (from \SI{1.6}{\kelvin} to \SI{25}{\kelvin}) an absorbing surface (28\% of emissivity), in an integrating cavity coupled to a Winston horn (50\% of efficiency). The lamp is placed at the center of the Lyot stop of the OLIMPO reimaging system and has an aperture which fills the central 5\% of the Lyot stop area.

\begin{figure}[htbp]
\begin{center}
\includegraphics[scale=0.31]{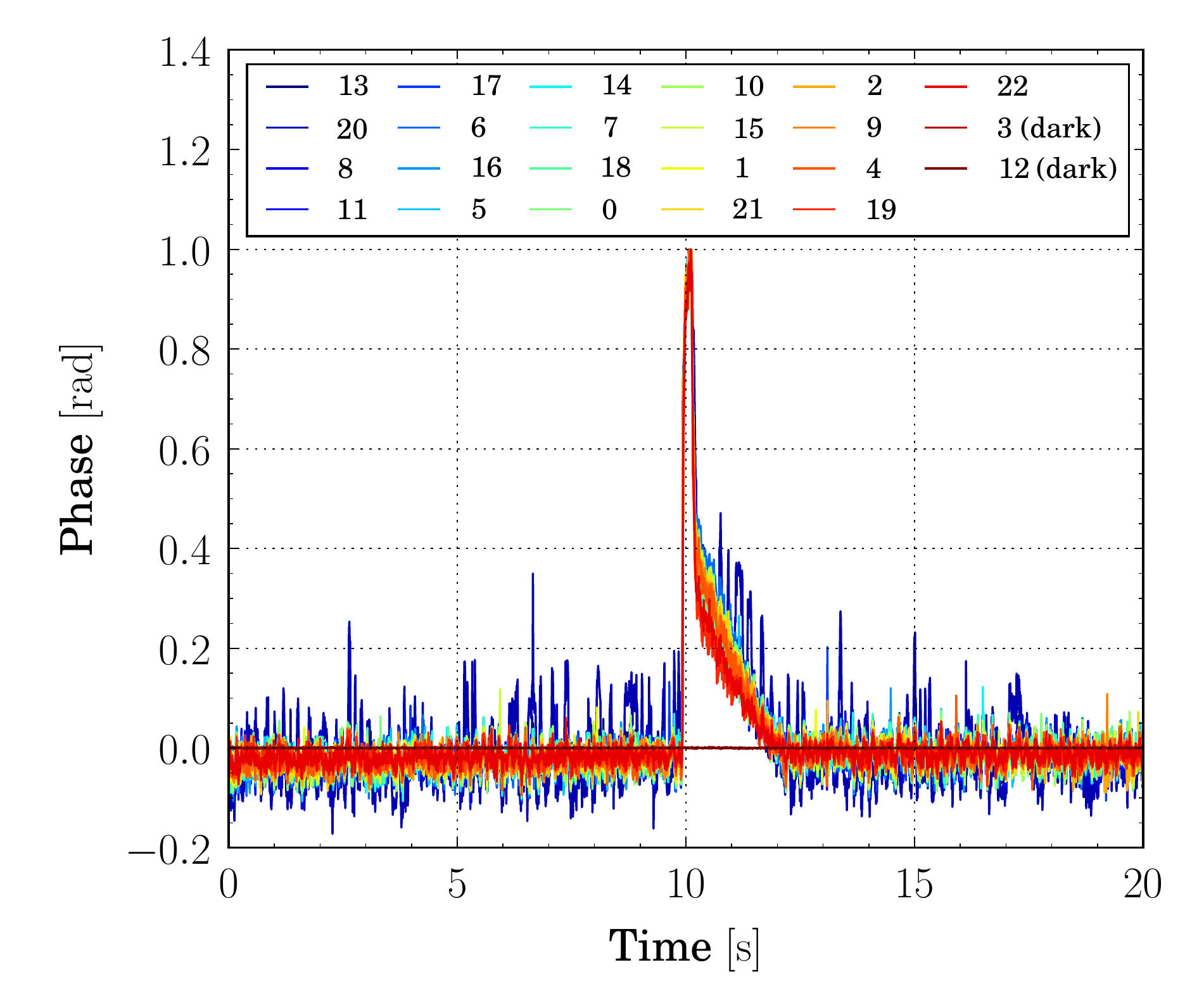}
\includegraphics[scale=0.31]{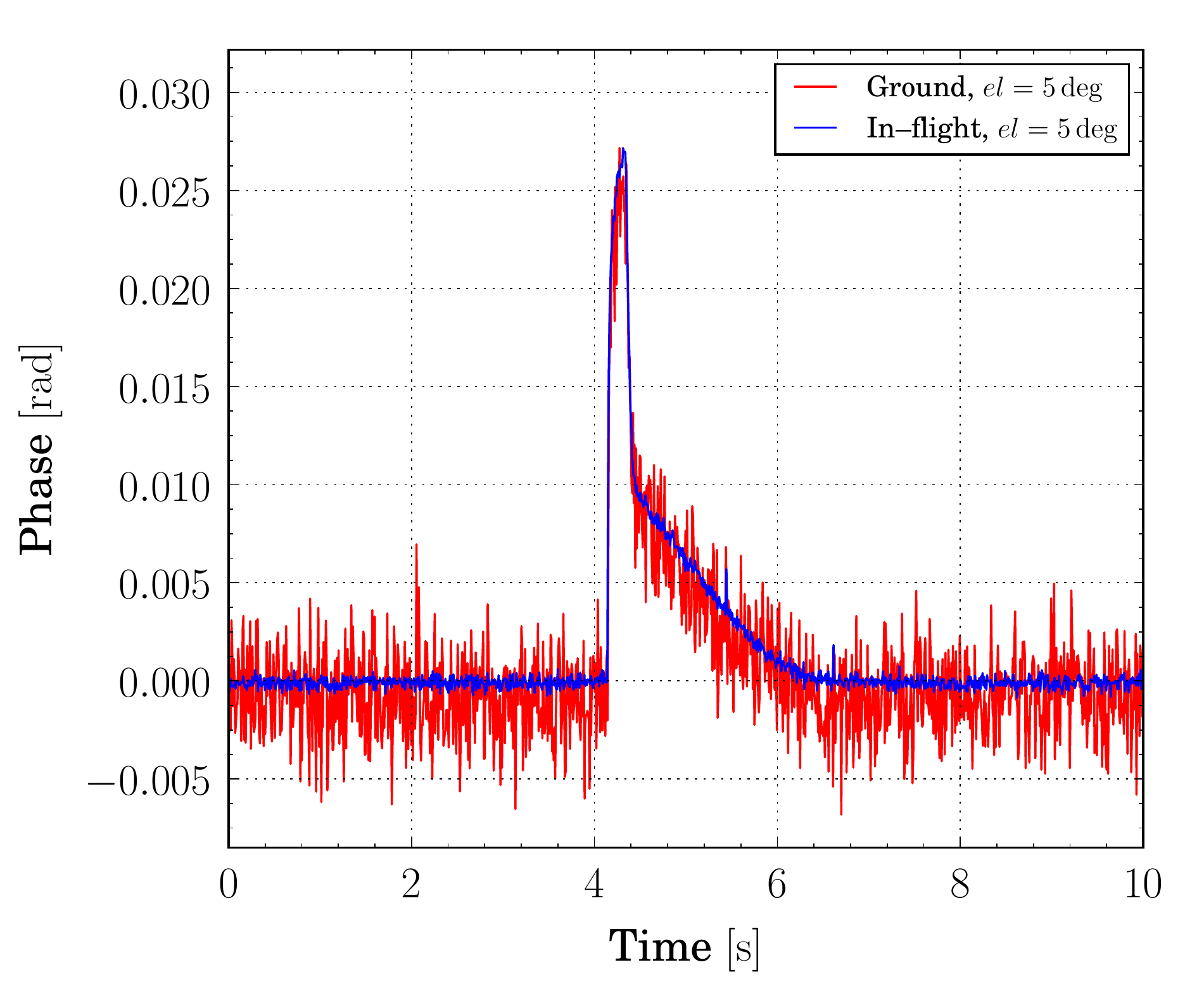}
\caption{Calibration lamp signals. {\it Left panel}: superposition of the signals detected by the pixels of the \SI{350}{\giga\hertz} array. {\it Right panel}: comparison between the signals at ground ({\it red}) and in--flight ({\it blue}), for pixel \#22 of the \SI{350}{\giga\hertz} array. The observed tails are due to the modulation of the calibration lamp current, which consists in a initial sharp pulse followed by a lower level linear decay.}
\label{fig:callamps}
\end{center}
\end{figure}

As a representative example, the {\it left panel} of fig.~\ref{fig:callamps} shows the superposition of the calibration lamp signal (normalized to \SI{1}{rad}) on the detectors of the \SI{350}{\giga\hertz} array from the noisiest to the least noisy. As expected, the signal is not detected by the two dark pixels. The {\it right panel} shows the comparison between the calibration lamp signal on the ground and in--flight for pixel \#22 of the \SI{350}{\giga\hertz} array. Both signals are normalized to the same value in order to preserve all the information in the noise.

This comparison for all the pixels of the arrays, allows to obtain the in--flight performance in terms of noise equivalent power (NEP), collected in tab.~\ref{tab:performance}, from which results that the performance of the \SI{250}{\giga\hertz}, \SI{350}{\giga\hertz} and \SI{460}{\giga\hertz} arrays is photon--noise limited, while that of the \SI{150}{\giga\hertz} array is very close to be photon--noise limited.

\begin{table}[htb]
\vspace{-0.4cm}
	\centering
		\fontsize{0.28cm}{0.45cm}\selectfont{
		\begin{tabular}{c|c|c|C|C}
		\hline
		\hline
		\multicolumn{1}{c|}{\multirow{1}{*}{Channel}}&
		\multicolumn{1}{c|}{\multirow{1}{*}{FWHM}}&
		\multicolumn{1}{c|}{\multirow{1}{*}{photon--noise--limited}}&
		\multicolumn{2}{c}{\multirow{1}{*}{average NEP $[\SI{}{\atto\watt/\sqrt{\hertz}}]$}}\\
		\cline{4-5}
		\multicolumn{1}{c|}{\multirow{1}{*}{$\left[\SI{}{\giga\hertz}\right]$}}&
		\multicolumn{1}{c|}{\multirow{1}{*}{$\left[\SI{}{\giga\hertz}\right]$}}&
		\multicolumn{1}{c|}{\multirow{1}{*}{NEP $[\SI{}{\atto\watt/\sqrt{\hertz}}]$}}&
		\multicolumn{1}{C|}{\multirow{1}{*}{ground}}&
		\multicolumn{1}{C}{\multirow{1}{*}{in--flight}}\\
		\hline
		\hline
150&25&68&$178\pm 27$&$89\pm18$\\
250&90&105&$879\pm 132$&$109\pm 22$\\
350&30&94& $289\pm 43$&$83\pm17$\\
460&60&211& $771\pm 116$&$178\pm 36$\\
\hline
		\hline
		\end{tabular}		
		}
		\caption{Array--averages of the NEP measured at ground and in--flight, compared to the photon--noise--limited NEP, expected in--flight. The values of the full width half maximum (FWHM) of the spectral band of each array are reported (measured in \cite{Paiella_ground}).}
	\phantomsection\label{tab:performance}
	\vspace{-0.4cm}
\end{table}

\subsection{Data contamination by cosmic rays}
\label{subsec:cosmic}

Estimating the fraction of data contaminated by cosmic rays (CRs) hitting the arrays is important to assess the susceptibility of KID technology to the near--space environment. It is therefore relevant for TRL advancement in view of future space missions. 

The analysis, described in detail in \cite{Masi_inflight}, takes into account also the contamination due to CRs embedded in the noise, but still affecting the sensitivity of the receiver \citep{Masi_cosmic}. Our approach consists therefore in histogramming the data and fitting them with the superposition of a model for the CR distribution and Gaussian noise. The model for the CR amplitude distribution is given by 

$$
\frac{dN}{dS} =2 \pi  A_o \frac{dN}{dA d\Omega} \frac{\left(k x\right)^2}{S^3}\;, 
$$
where $A_o$ is the area of the Si wafer where the pixels are deposited, $dN/dA/d\Omega$ is the number of CRs hitting the wafer per unit area and solid angle, $x$ is the substrate thickness, $S$ is the signal amplitude produced by the CR, and $k$ is a proportionality constant which takes into account the responsivity and the energy deposited by CRs in the wafer.

\begin{figure}[htbp]
\begin{center}
\includegraphics[scale=0.35]{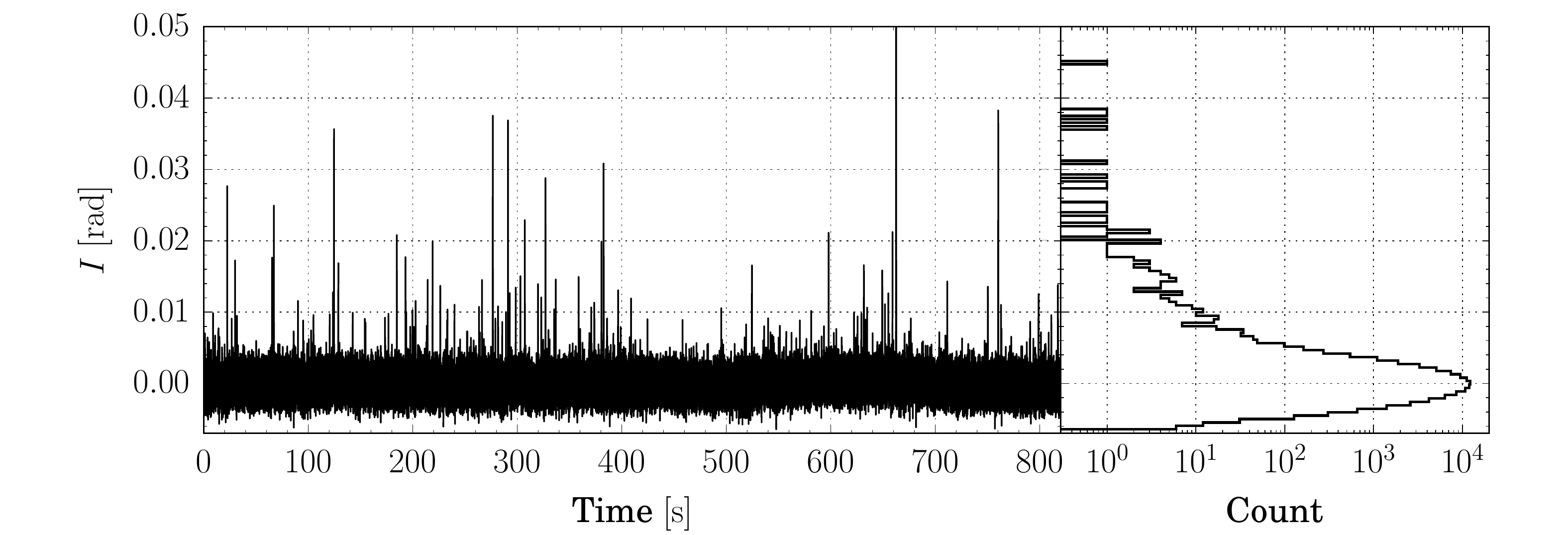}
\includegraphics[scale=0.35]{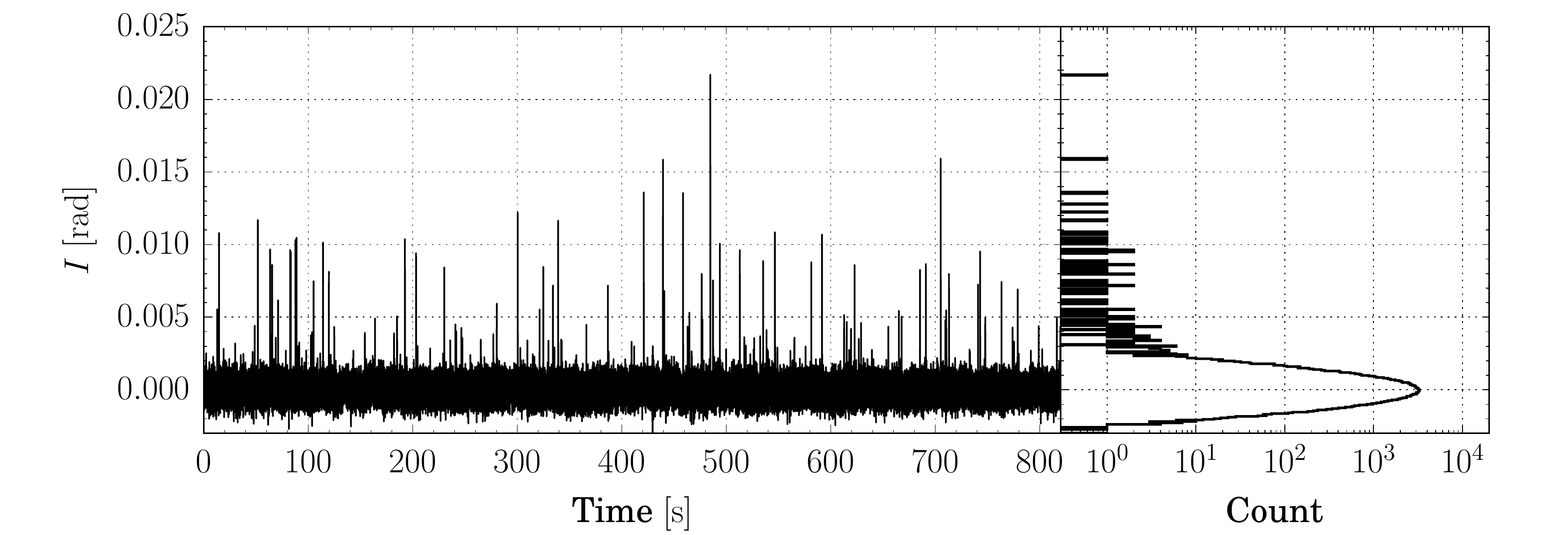}

\caption{\SI{820}{\second}--long timestream ({\it left}) and corresponding histogram ({\it right}) for the center pixel of the \SI{150}{\giga\hertz} ({\it top}) and the \SI{460}{\giga\hertz} ({\it bottom}) array.}
\label{fig:CR}
\end{center}
\vspace{-0.2cm}
\end{figure}

As a representative example, the {\it left panels} of fig.~\ref{fig:CR} show a \SI{820}{\second}--long timestream (\SI{100000}{} samples) for the center pixel of the\SI{150}{\giga\hertz} ({\it top}) and \SI{460}{\giga\hertz} ({\it bottom}) array, while the {\it right panels} show the corresponding histogram. In this period the telescope boresight was stable and the detectors were tuned. Fitting the histograms and integrating the best fit of the CR amplitude distributions between the minimum and the maximum amplitude, we estimated the fraction of contaminated data \cite{Masi_inflight}. The results are collected in tab.~\ref{tab:CR}.    

\begin{table}[htb]
\vspace{-0.3cm}
	\centering
		\fontsize{0.28cm}{0.45cm}\selectfont{
		\begin{tabular}{c|c}
		\hline
		\hline
		\multicolumn{1}{c|}{\multirow{1}{*}{Channel}}&
		\multicolumn{1}{c}{\multirow{1}{*}{Average limits on the}}\\
		\multicolumn{1}{c|}{\multirow{1}{*}{$\left[\SI{}{\giga\hertz}\right]$}}&
		\multicolumn{1}{c}{\multirow{1}{*}{fraction of contaminated data}}\\
		\hline
		\hline
150&$<2.7\%$\\
250&$<2.8\%$\\
350&$<0.1\%$\\
460&$<0.2\%$\\
\hline
		\hline
		\end{tabular}		
		}
		\caption{Array--average limits on the fraction of contaminated data by CRs.}
	\phantomsection\label{tab:CR}
	\vspace{-0.4cm}
\end{table}

These results on the impact of CRs, obtained for the OLIMPO LEKIDs at \SI{37.8}{\kilo\metre} of altitude, can be considered, as a first approximation, representative of a low--Earth orbit satellite mission, and pave the way to the use of KID technology in satellite missions.

\section{Conclusion}
\label{sec:conclusion}
This paper summarizes the technical results on the detectors obtained during the first hours of the OLIMPO flight. They show that the detectors can be tuned in--flight, and their performance is significantly improved with respect to ground, due to reduced radiative background, more stable conditions, and reduced microphonics. We obtained photon--noise--limited performance for the channels at 250, 350 and \SI{460}{\giga\hertz}, and very close to photon--noise--limited performance for the \SI{150}{\giga\hertz} channel. We find that the contamination of data due to cosmic ray events is less than about 3\% for all the pixels of the arrays.

These results represent an important step in the TRL advancement of KID technology for space, in view of future satellite missions.

\begin{acknowledgements}
This activity has been supported by the Italian Space Agency.
\end{acknowledgements}

\pagebreak

\end{document}